# High-Percentage Success Method for Preparing and Pre-Evaluating Tungsten Tips for Atomic-Resolution Scanning Tunneling Microscopy

J.K. Schoelz, P. Xu, S.D. Barber, D. Qi, M.L. Ackerman, G. Basnet, C.T. Cook, and P.M. Thibado[a), b)]

Department of Physics, University of Arkansas, Fayetteville, Arkansas 72701

[a)]American Vacuum Society member.
[b)]Electronic mail: thibado@uark.edu

A custom double-lamella method is presented for electrochemically etching tungsten wire for use as tips in scanning tunneling microscopy (STM). For comparison, tips were also manufactured in-house using numerous conventional methods and examined using an optical microscope. Both sets of tips were used to obtain STM images of highly-oriented pyrolytic graphite, the quality of which varied. The clarity of the STM images was found to be correlated to the optically-measured cone angle of the STM tip, with larger cone angles consistently producing atomically resolved images. The custom etching procedure described allows one to create larger cone angles and consequently proved superior in reliably producing high-quality tips.

# I. INTRODUCTION

Reliably fabricating ultrasharp metallic probes with a tip apex radius on the order of 10 nm has challenged researchers since the debut of the field electron emission microscope by Müller in 1936, and has become even more significant as field ion microscopy and scanning tunneling microscopy (STM) have gained prominence and remained vital research techniques. Because the tip apex radius is such a critically important attribute when attempting to image on the atomic scale, experiments characterizing ultrasharp probes have primarily focused on the smallest scales, employing scanning electron microscopy (SEM) which is capable of 100,000× magnification. Such studies have established that ultrasharp metallic probes can, in fact, be reliably manufactured by electrochemical etching, which is a simple, inexpensive, and widely used technique.[1] For STM, tungsten wire is preferred because of its high conductivity, mechanical strength, durability, and low cost.[2] Typically, electrochemical fabrication methods involve submerging the tungsten wire and a conducting ring into an electrolytic conducting solution, then applying a bias voltage of either DC[3] or AC[4] between them. The resulting current between the ring and wire (mediated by the solution) drives a reduction-oxidation reaction, which oxidizes the tungsten at the air/solution interface. This basic scheme allows for many variations, each with its own advantages. A great deal of work has thus been devoted to developing optimal etching procedures for consistently and feasibly producing high-quality tips.[5]

One of the most recently developed tip fabrication techniques is to use tungsten wire arranged horizontally under a high-magnification optical microscope, with the loop attached to a micromanipulator for fine motion control.[6] The solution forms a lamella suspended in the loop, which is moved back and forth over the tungsten wire while etching to create a tip of the desired shape. This method, known as zone electropolishing, offers superior control and precision, and even allows for re-etching a damaged or crashed STM tip. However, this method does pose a problem in the final drop off step. It's necessary to thin a small section near the end of the wire into a "neck" shape, and then precisely sever the wire while cleanly separating the extra end piece. Finally, when the extra piece is removed, the etching must stop immediately to avoid detrimental back etching, which quickly dulls the tip. The success of this technique is dependent upon the patience, skill, and reaction time of the technician. To address these potential difficulties, automatic etching systems have been constructed which monitor the current flowing through the etchant and use feedback circuitry to terminate power immediately upon completion of the tip etching (typical electrical cutoff time is on the order of 10 ns). Unlike the micromanipulator method, in these systems the tungsten wire is oriented vertically, and all components remain stationary during etching. Early designs involved submerging the loop and wire in the electrolyte solution for the duration of the etching, which provided unmatched simplicity. However, Klein and Schwitzgebel reported an automatic method in 1997 that offered a far greater degree of control over the final tip shape by using a lamella rather than submersion.[7] Nevertheless, cutoff circuits, in practice, often stop the etching too soon due to natural fluctuations in the current, making it necessary for the operator to constantly monitor the process and occasionally restart it. In addition, the change in current upon completion of the etching is occasionally too small to trigger the cutoff circuit, resulting in the back etching problem. A novel alternative is the mechanical or gravity switch developed by Kulawik *et al.* in 2003 which utilizes two lamellas to break the etching circuit as soon as the tip is finished etching.[8] In this setup, the current flows through the end of the wire that is being etched off. Naturally, once etched through, the lower part of the wire drops under the influence of gravity. This causes circuit breaking when the wire breaks contact with the solution held in the upper lamella, providing a reliable cutoff time of about 1-10 ms, depending on the thickness of the suspended fluid.

In this note, we present a custom electrochemical etching procedure which incorporates the best features of the common methods just discussed. Our double-lamella system reliably produces ultrasharp tungsten probes capable of producing STM images with atomic resolution, as demonstrated by testing the resulting tips on highly-oriented pyrolytic graphite (HOPG). STM tips were also manufactured using several other electrochemical techniques and were similarly tested for quality by using them to image HOPG on the atomic level. Furthermore, magnified optical images of the tips were acquired before they were transferred into the STM chamber. A strong correlation was found to exist between a tip's cone angle and its ability to produce images with atomic resolution. We propose that this observation is related to mechanical stability and can be used as a quick and economical test to evaluate the probable quality of a tip, assuming it was etched by a typical electrochemical method that has been shown by SEM to consistently yield sufficient sharpness at the apex.

## II. EXPERIMENTAL

For comparison and to create additional tips with different characteristics, the popular horizontal zone electropolishing method, the simple submersion method, a single-lamella vertical method, and our double-lamella method were employed. The zone electropolishing method was used most often. This technique requires AC voltage, high magnification, and with a horizontal tungsten wire. In the submersion method, a single gold ring with a diameter of a few centimeters was submerged into the NaOH solution, and a DC bias ranging from +3.0 to +6.0 V was applied to it. The tungsten wire was then lowered until only 2-3 mm remained above the surface of the solution and 2 mm was below the surface. The etching rate along the wire decreases quickly as the distance below the air/solution interface increases, resulting in an atomically sharp tip. A differential cutoff circuit (Omicron Tip Etching Control Unit) was used to automatically discontinue the bias and stop the etching when the current experienced a sharp drop (i.e., when the lower part of the tungsten wire broke off and dropped into the solution).[9] The single-lamella technique was very similar to the submersion technique; although the etching film is thinner and therefore produces a larger cone angle tip. A differential cutoff circuit was again used to automatically discontinue the bias when the bottom part of the wire fell.

For this study, over two hundred STM tips were produced from 0.25-mm diameter polycrystalline tungsten wire. Our final, optimized setup was a vertically oriented, double-lamella, gravity switch system, as shown in Fig. 1(a). The etching process is powered by a Keithley 2400 Sourcemeter, as shown on the right side of the photograph. This instrument supplies a constant DC voltage throughout the etching process and also displays the current flowing through the circuit. The eye pieces for the 30× magnification microscope can be seen in the lower central region. Rather than being upright according to the original design, the optical microscope is on its side (nearly horizontal), mounted on a custom support in a position that allows the tungsten wire to be oriented vertically during the etching progress. The STM tip is held at a fixed position equal to the focal position of the microscope using a magnetic stand which is positioned to the right of the microscope, as also shown in Fig. 1(a). The magnification provided by the microscope also facilitates careful regulation of the thicknesses and positions of the lamellae. The position of the two loops is controlled using a micromanipulator, which offers coarse adjustment in the x, y, and z directions, as well as a fine hydraulic control for the z direction in order to maintain etching at the desired location with minimal vibration. The micromanipulator can be seen to the left of the microscope near the top of the photograph. The precision movement of the loops is necessary because zone etching changes the shape of the wire, causing the top lamella (where etching occurs) to shift and adhere to a slightly different site. In addition, the top lamella may pop several times before the tip is finished etching, and requiring rewetting, followed by thinning. Note that it is necessary to readjust the position of the loops to form around the same point on the wire as previously.

A zoomed-in photograph of the tip, tip holder, and gold loops during the etching process is shown in Fig. 1(b). The etching process begins with the insertion of a 5 mm length of 0.15-mm diameter tantalum (Ta) wire into a stainless steel Omicron tip holder, followed by a 10-mm length of the tungsten (W) wire. The tantalum is relatively soft and as it deforms, it serves to hold the tungsten in place as it is pressed into the tip holder. The tantalum wire is then trimmed to a length of a couple millimeters and bent back so as not to interfere with the STM operation. The advantage of the Ta wire press-in method becomes apparent when the tip needs replacement, as it may be removed along with the used tungsten tip, and the holder reused. A magnet keeps the tip holder in place during the etching process. Because the tungsten wire is fixed at the focal length of the microscope, the two gold loops are raised up to surround the tungsten wire until the upper gold loop is within about 2-3 mm of the end of the Omicron tip holder, the maximum length to avoid damage during transfer to the scanner assembly inside the STM chamber after the tip is complete. The full etching circuit and the mechanism behind the gravity switch are also illustrated in Fig. 1(c). The upper loop is attached to the positive side of the DC supply, while the lower loop is attached to the grounded side (this moves the gas formation away from the etching site). When the wire is etched away, the lower section of the wire drops due to gravity. This action results in the electrical etching circuit being broken as soon as the falling wire separates from the etching solution suspended in the upper loop (falling time is about 10 ms for a 1 mm thick lamella). The fall stops the etching process, even though the tip is still submerged in the etching solution contained in the upper loop. This method works well and eliminates the need to rely upon human intervention or a special response characteristic within the electrical circuit.

A magnified view of the tantalum wire, tungsten wire and two gold loops is shown schematically in Fig. 1(d). The two gold rings are separated by about 5 mm and oriented so that their faces are parallel and lie in horizontal planes. The top ring is 15 mm in diameter, while the lower ring has a diameter of about 5 mm. A beaker containing a solution of 8g NaOH dissolved in 100 mL of deionized water is raised to briefly submerge the rings. When the beaker is lowered, a thin film of the conducting solution is left suspended across each ring. With the tungsten wire in place, a meniscus forms around the wire at each ring from the suspended solutions as indicated in Fig. 1(d). A thick meniscus was observed at the top ring, where the wire was etched, resulting in a longer length of the tungsten wire etching tip, which, in turn, creates a smaller cone angle. To achieve a thinner meniscus and therefore, a larger cone angle, some of the suspended solution can be carefully wicked away. Further, it is important to monitor the meniscus and move it down the tapered section of the wire as it will attempt to climb up the wire. Another, important factor is to make sure the tip is mounted nearly vertical, so during the final etch step the wire will not rotate and tear the end of the W tip.  During etching, the power supply was set to apply a DC bias of about 8 V during the tip making process.  A higher voltage results is faster etching. For our setup an 8 V setting generated a potential difference of about 4.5 V between the tungsten wire and upper gold loop.  Note, that the upper loop is held at a positive bias so that the bubble formation happens at the lower loop.  This stops the bubbles from interfering with the etching process and also allows for a clear view of the tip throughout the etching.  Also, as the top wire is thinned the meniscus favors climbing up the W wire and it is important to monitor this effect and move the loop down so the solution does not etch the wire above its original starting location.

At the conclusion of each described method, the completed STM tip was thoroughly rinsed with distilled water, then isopropanol, and finally swirled in a concentrated HF solution for 30 seconds to remove any tungsten oxides. The tips were then placed under an optical microscope and photographed under 100×, 500×, and 1000× magnification before being immediately transferred through a load-lock into the STM chamber.

Experimental STM images were obtained using an Omicron ultrahigh-vacuum (base pressure of $10^{-10}$ Torr), low-temperature STM operated at room temperature. The top layers of a 6 mm × 12 mm by 2 mm thick piece of HOPG were exfoliated with tape to expose a fresh surface. It was then mounted with silver paint onto a flat tantalum STM sample plate and transferred into the STM chamber. Large-scale (100 nm × 100 nm) and small-scale (6 nm × 6 nm) filled-state images of the sample were obtained with all the tips, using a tunneling current of 0.1 nA and a tip bias of 0.1 V.

## III. RESULTS

Of the four techniques used to manufacture STM tips, we found that the horizontal zone electropolishing (approx. 70 tips) and the simple submersion (approx. 80 tips) methods mostly produced tips with small cone angles (< 10º). The single lamella (approx. 30 tips) and double lamella (approx. 30 tips) methods mostly produced tips with larger cone angles (> 10º). In general, the fabricated STM tips were sorted into three broad categories according to the STM image quality each was able to obtain shortly after tunneling (i.e., directly after the approach and without any voltage pulses or other tip cleaning procedures). The best tips were defined as those which produced small-scale images in which the individual atoms of the HOPG surface were clearly resolved. Medium quality tips were defined as those producing small-scale images where atomic features, such as atomic rows, were resolved rather than individual atoms. The lowest quality tips were those for which no atomic resolution was displayed in the small-scale images, but which did resolve monolayer steps of graphite in large-scale images. An example of each category of STM image quality is shown on the left side of Fig. 2. The quality of the images increases going up the vertical axis, beginning with a large-scale, low-quality image on bottom, a small-scale, medium-quality image in the middle, and a small-scale, high-quality image on top. Note each image was minimally processed, including a plane subtraction and minor filtering. In addition to the classification of HOPG images, the optical images of the STM tips themselves were also reviewed. Their physical characteristics were documented, and a strong correlation was noted between the cone angle of the tips, as measured in the optical images, and the quality of the STM image obtained using that tip. These results are also summarized in Fig. 2. The cone angle of a tip was defined as the full angle between the two perimeter lines of the cone, as illustrated in the bottom-left corner of Fig. 2. This shows an ideal atomically sharp yet stable tip having a full cone angle of about 60º and is generated by stacking spheres into a pyramid. To the right of the illustration and along the bottom of the figure, examples of the optical images are displayed. The cone angle increases going along the horizontal axis with small (0-10°), medium (10-15°), and large (>15°) cone angles. Note, the large cone tip displayed was our largest and best (cone angles around 20º were more typical). The entries in the resulting 3×3 matrix are the approximate percentages of tips which produced the given quality of STM image for each type of cone angle. Note that the percentages in any row or column sum to 100%. The table can be read left to right as well as top to bottom. For example, the top row represents 100% of all the high quality STM images. The first cell indicated that 10% of the high-quality images came from STM tips that have a small cone angle, another 20% can from tips with medium cone angles, and most importantly, 70% of the tips that provided high-quality STM images had a large cone angle. This information is repeated in the second row for the medium quality STM images and the in the third row for the lowest quality STM images. Notice that for the lowest quality images, the vast majority came from the STM tips that had a very small cone angle. The matrix can also be read by looking at the individual columns. The left column corresponds to 100% of the tips with small cone angles (approx. 100 tips); 70% of these tips generated low quality STM images, while 20% generated medium quality images, and 10% generated high quality images. The middle column corresponds to tips with medium cone angles (approx. 60 tips); 60% of which generated medium quality images, while 20% generated low quality images, and 20% generated high quality images. The last column corresponds to tips with large cone angles (approx. 50 tips); 70% of which generated high quality images, while 20% generated medium quality images, and 10% generated low quality images. In general, we found tips with a cone angle of ~15° or greater gave excellent STM images. We believe a cone angle of ~60° would be optimal for both stability and sharpness as shown schematically in Fig. 2.

## IV. DISCUSSION

We found the results summarized in Fig. 2 to be very surprising. From our findings, it is clear that an optically-measured cone angle of an STM tip is the single greatest factor in determining the quality of the images obtained using that tip. By looking at this large scale optical property of the STM tip, the mechanical stability of the STM tip may be indirectly observed. To substantiate this observation, we made an STM tip by submerging a slightly longer tungsten wire into the electrochemical solution (i.e., 3-4 mm instead 2 mm), which resulted in a long, thin tip. Under the high-magnification optical microscope we observed the completed tip spontaneously vibrating with an amplitude of about one micron. If an STM tip vibrates during the scanning process involved in data taking, then each data point in the resulting STM image is a spatial average over a length scale similar to the amplitude of these vibrations, resulting in the poorest quality images.

Naturally, it is critical that an STM tip be atomically sharp at its apex. SEM studies confirm that electrochemical etching generally produces ultrasharp tips. Thus, within the scope of the best electrochemical etching techniques, it is important to further characterize the STM tips using a simple optical microscope to ensure that they are of the highest quality.

In this study, the various lamella techniques gave the greatest control over the size of the cone angle. The popular horizontal zone electropolishing method was shown to primarily produce less useful small cone angle STM tips. The reason for this is that the loop is moved back and forth along the tungsten wire in this method, so the wire is etched over a longer length, creating a smaller cone angle. Nevertheless, the zone electropolishing method does offer some advantages. This study brings together the best features of the various methods, as shown in Fig. 1, to consistently produce STM tips which yield a high percentage of atomic-resolution images. One of the factors that most influenced our choices was the risk of back etching, which was a primary reason for selecting the double-lamella gravity switch approach. Additional improvements were made, however, within this framework. The second important factor under consideration was our ability to control the shape of the tip, especially its cone angle, as much as possible. Using a lamella to etch the wire gave a degree of control over the final tip shape that was impossible to obtain in a submersion method. Having an optical microscope focused on the tip combined with the manipulator's control allowed us to observe and modify the position of the lamella, making it possible to predict the cone angle of the finished tip. A positive DC etching voltage was chosen to eliminate disruptive gas formation at the etching site. The result of all of these choices was a reliable fabrication method that generated, with a 70% success rate, STM tips capable of atomic resolution.

## V. CONCLUSIONS

In conclusion, STM tips were electrochemically etched using an in-house developed, double-lamella method which offers excellent control over the tip's shape and consistently produces ultrasharp probes capable of atomic-resolution STM images. The setup has the tungsten wire oriented vertically while under optical magnification and uses a manipulator to tune the process as it evolves. The resulting tips were photographed under 1000× magnification and then used to acquire STM images of HOPG in order to determine their overall quality. A strong positive correlation exists between the cone angle of a tip, as determined by the optical images, and the chance that the tip will be capable of atomically resolved STM images. Larger cone angle tips are more likely to provide better STM images, as a result of their greater mechanical stability. Optical imaging of the tips can be used to accurately predict their quality, making it an easy and inexpensive tool for evaluating electrochemically etched tips prior to STM imaging.

## ACKNOWLEDGMENTS

This work is supported in part by the National Science Foundation (NSF) under grant number DMR-0855358 and the Office of Naval Research (ONR) under grant number N00014-10-1-0181.

**Figure Captions**

Figure 1. (a) A photograph of the entire tip etching setup. The tip wire is held fixed at the focal length of the microscope using a magnetic support. Two gold loops are mounted on a micromanipulator with course x, y and z control which is located to the left. A hydraulic fine z control is used to alter the etching position. A DC voltage is applied between the two gold loops using a Keithley 2400 Sourcemeter shown to the right. (b) A zoomed-in view of the tungsten wire mounted inside the tip holder and held in position to be etched. The two gold loops are electrically connected by the tungsten wire. (c) Schematic of the double lamella setup showing the tungsten wire inserted into the tip

holder with a short length of tantalum wire to hold it in place.  The tip holder is held in place using a magnet.  The DC power supply wiring results in the top loop acting as the cathode and the bottom loop as the anode. NaOH solution is suspended in each loop creating an electrochemical cell in which the etching occurs in the upper loop.
(d) Schematic of the local etching setup showing the two gold loops. The width and location of the NaOH suspended fluid determines the cone angle of the resulting tip.

Figure 2. Chart relating the cone angle of the etched tip to the probability that the tip will be capable of producing a certain quality STM image. In the bottom left corner is a schematic of what is imagined to be the ideal stacking pattern for sharp, stable STM tip (Note, the stack gives a cone angle of 60º). Along the horizontal axis are three cone angle ranges along with representative photos of STM tips in the cone angle range (images were taken at a magnification of 1000×).  Along the vertical axis are three STM images taken with a tunneling current of 0.1 nA and a tip bias of 0.1 V of the HOPG surface.  The bottom image is a 100 nm x 100 nm image.  A monolayer step is visible on the surface.  Above this is a 6 nm x 6 nm image of HOPG.  In this image, it is possible to resolve rows of atoms on the HOPG surface but atomic resolution could not been achieved.  The top image is a 6 nm x 6 nm image of the HOPG surface with atomic resolution.

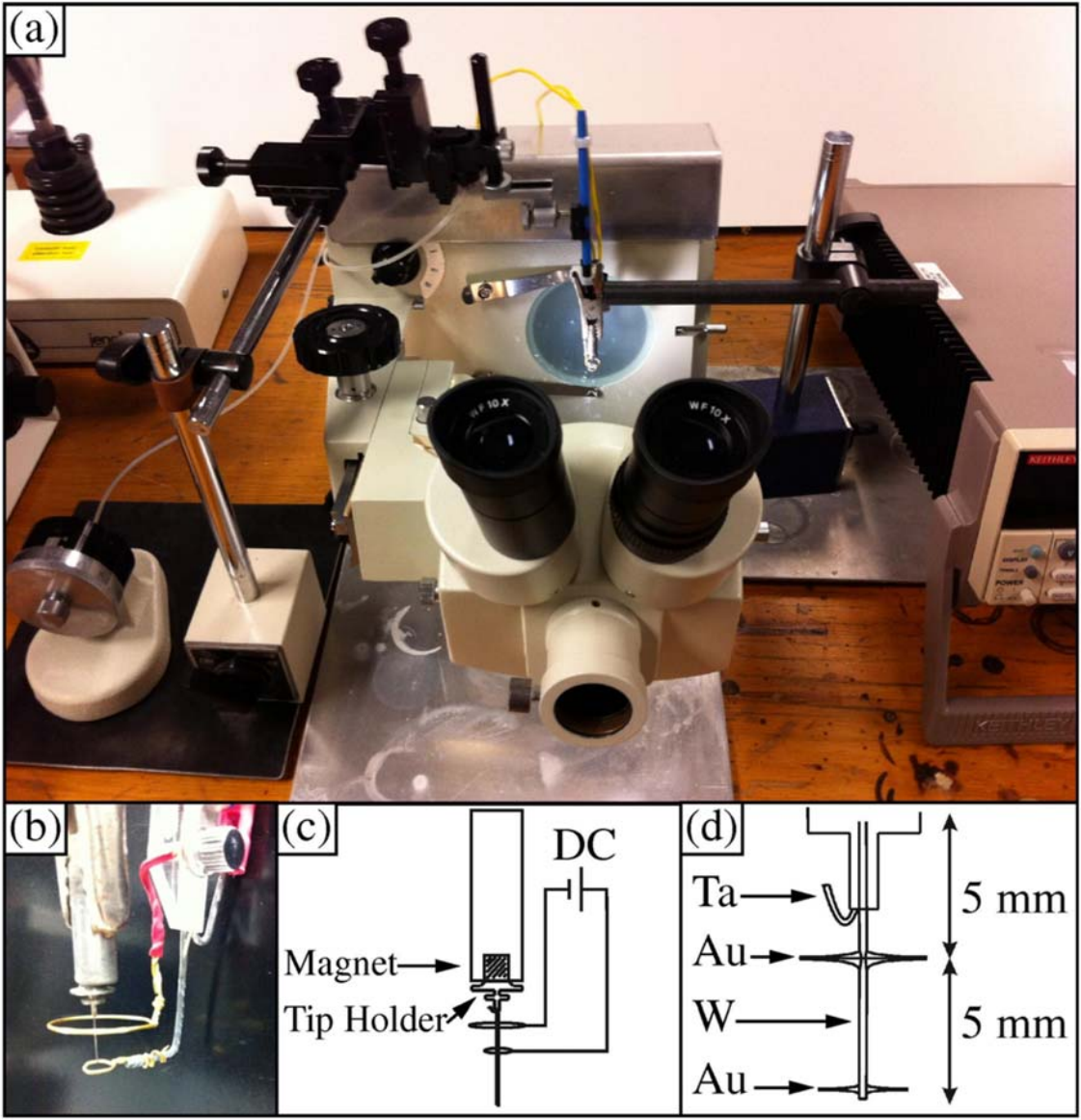

Figure 1

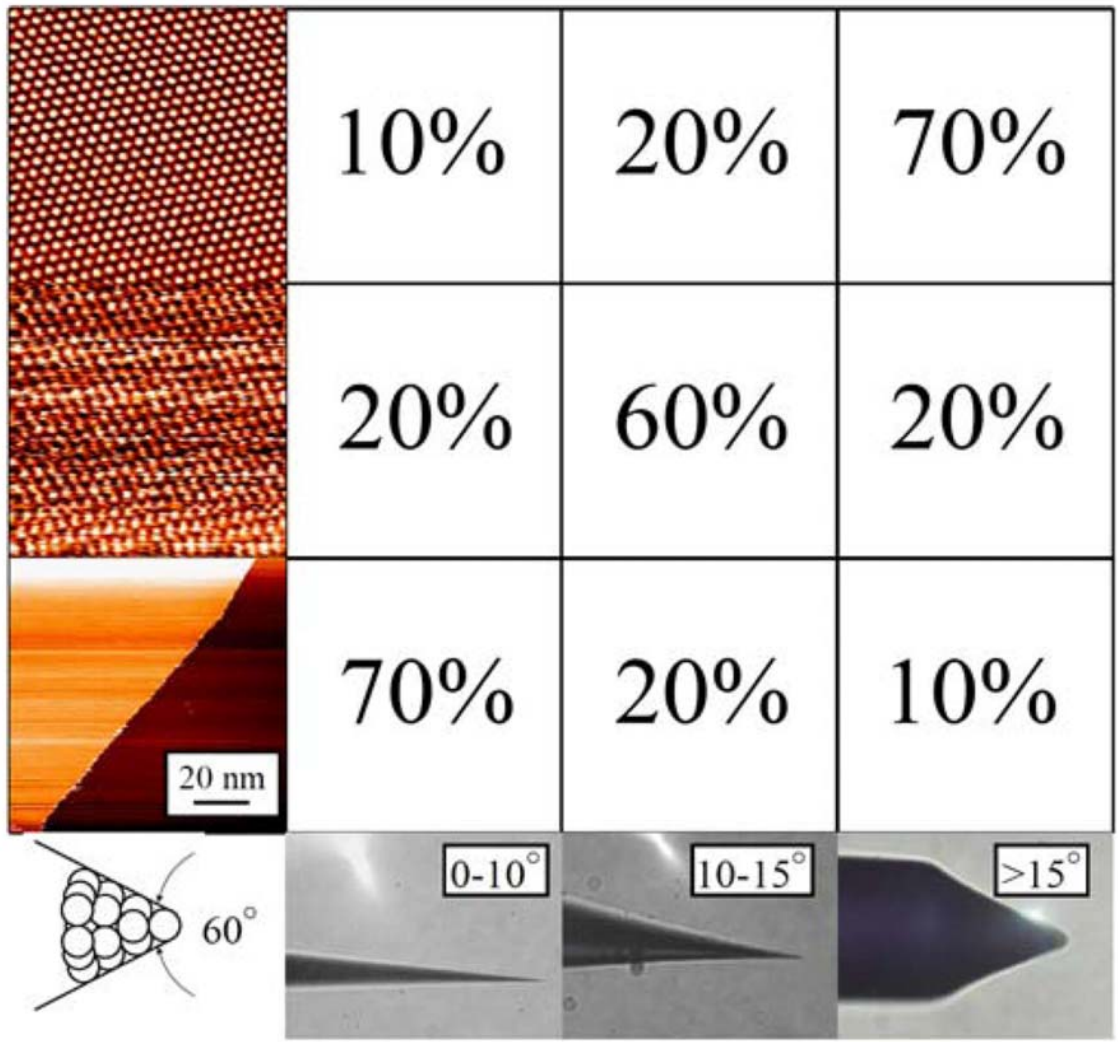

Figure 2